\documentclass[usenatbib]{mn2e}
\usepackage[fleqn,reqno]{amsmath}
\usepackage{graphicx}
\usepackage{txfonts}
\def\hh{H$_{\rm 2}$}

\def\ammo{NH$_3$}

\def\nddh{ND$_2$H}

\newcommand{\jcp}{J.Chem.Phys.}

\def\apj{ApJ}

\def\aap{A\&A}

\begin{document}
   \title{Collisional excitation of doubly deuterated ammonia \nddh ~
     by para-\hh}

   \author[L. Wiesenfeld, E. Scifoni, A. Faure,  and E. Roueff]{L. Wiesenfeld $^1$, E. Scifoni $^{1,2}$, A. Faure $^1$,  and E. Roueff $^3$\\
 $^1$ Laboratoire d'Astrophysique de Grenoble, CNRS/Universit\'e Joseph-Fourier, Grenoble, France.\\
 $^2$ Department of Biophysics, Gesellschaft f\"ur Schwerionenforschung, Darmstadt, Germany.  \\
$^3$ LUTh, CNRS/Observatoire de Paris, Meudon, France
}

   \date{\today}
  \maketitle 
  \begin{abstract}

   Collisional de-excitation rates of partially deuterated molecules
   are different from the fully hydrogenated species because of
   lowering of symmetry. We compute the collisional (de)excitation
   rates of \nddh ~ by ground state para-\hh, extending the previous
   results for Helium. We describe the changes in the potential energy
   surface of \ammo - \hh ~involved by the presence of two deuterium
   nuclei. Cross sections are calculated within the full
   close-coupling approach and augmented with coupled-state
   calculations. Collisional rate coefficients are given between 5 and
   35~K, a range of temperatures which is relevant to cold
   interstellar conditions. We find that the collisional rates of
   \nddh ~ by \hh ~ are about one order of magnitude higher than those
   obtained with Helium as perturber. These results are essential to radiative transfer modelling
   and will allow to
   interpret the millimeter and submillimeter detections of ND$_2$H
   with better constraints than previously.
\end{abstract}

%
%________________________________________________________________

\section{Introduction}

Doubly deuterated ammonia has been detected for the first time by
\citet{roueff:00} towards the dark cloud L~134N via its
$1_{10}-1_{01}$ ortho and para transitions at 110~GHz and subsequently
in the protostellar environment L1689N by \citet{loinard:01}.  The
sub-millimeter fundamental transitions have then been detected at the
Caltech Sub-millimeter Observatory by \citet{roueff:05, lis:06} and on
the APEX antenna in Chile by \citet{gerin:06} towards the Barnard 1
molecular cloud and L1689N. \nddh\ was shown to be a sensitive tracer
of the physical conditions of this star forming region. In a previous
paper \citep{machin:07} some of us have presented the collisional
formalism and computed the collisional excitation cross-sections and
rate coefficients of \nddh ~by He by introducing the appropriate
changes in the potential energy surface of \ammo ~- He, computed by
\citet{hod01}. This new accurate intermolecular potential energy
surface (PES) for NH$_3$-He has also been used previously to
reevaluate the collisional excitation of NH$_3$ and NH$_2$D by He
\citep{mac05,scifoni:07,mac06}.  Here, we present results for \nddh ~
in collision with para-\hh. H$_2$ is the main constituent of dark
clouds environment where \nddh ~ has been detected. Collisional
excitation is dominated by interaction with low temperature H$_2$;
this motivated our present study, aimed particularly at the low
temperature excitation schemes of \nddh.

The paper is organized as follows. In section \ref{sec:potential} we
briefly describe the changes introduced by the presence of two
deuterons in the PES of \ammo ~- \hh, published recently
\citep{maret:09} . We present in section \ref{sec:collision} the
collisional equations for the \nddh ~- \hh ~ system. The cross
sections and the corresponding reaction rate coefficients of the \nddh
~- \hh ~ system are given in section \ref{sec:result}. We present our
conclusions in Section \ref{sec:conclusion}
 
%__________________________________________________________________

\section{Potential energy surface}
\label{sec:potential}
We have recently put forward a potential energy surface (PES) of
NH$_3$ in interaction with H$_2$. This rigid rotor PES is reported in
full details in \citet{maret:09}. Here we need to transform this PES
from the triple H isotopologue of ammonia to the doubly deuterated
one, ND$_2$H. For a given \emph{inter}-molecular geometry, the
interaction between the molecules of ammonia and H$_2$ depends on the
various electronic charges (electrons and nuclei) as well as of the
\emph{intra}-molecular geometries of ammonia and molecular
hydrogen. For the two isotopologues NH$_3$ and ND$_2$H the charge
structure is identical, as long as we remain in the Born-Oppenheimer
approximation. However, the geometries are different in two aspects:
{\it (i)} the intra-molecular distances change slightly and {\it (ii)}
the position of the centre of mass of ND$_2$H is shifted with respect
to the centre of mass of NH$_3$ and the principal axes of inertia are
rotated. Let us treat both points in succession.

The original PES for NH$_3$-H$_2$ was computed with both monomer
geometries taken at their average value in the ground vibrational
state \citep{maret:09}.  This has been demonstrated to be a much
better approximation than taking equilibrium geometries: in the case
of the similar H$_2$O-H$_2$ system, \citet{alex05,valiron:08} have
found that employing state-averaged geometries is a very good
approximation for including vibrational effects within a rigid rotor
PES. We note that the importance of the zero-point vibrational correction was discussed earlier 
by \citet{meuwly:97}, in the $\rm N_2H^+ - He$ system.

In particular the $r$(HH) distance for molecular hydrogen was
shown to provide the largest vibrational correction. It was taken by
\citet{maret:09} at its average value of 1.4488~Bohr. Similarly, the
NH$_3$ molecule was taken at its ground-state average geometry with
the following parameter: $r(\mathrm{NH})= 1.9512$~Bohr,
$\widehat{\mathrm{HNH}}= 107.38$~deg. For ND$_2$H, the equilibrium
geometry is identical to NH$_3$ since the equilibrium bond lengths and
angles depend only on the electronic structure, within the
Born-Oppenheimer approximation. However the anharmonic averaging over
the ground state vibrational state changes slightly those values,
because of mass effects on the vibrational functions. For
H$_2$O/D$_2$O, these effects have been shown to be essentially
negligible with
respect to the shift of the center of mass \citep{scribano:10}.
 As a
result, the above ground-state averaged geometry of NH$_3$ can be
employed for any isotopologue of ammonia and was adopted here.

\begin{figure}
\centering \includegraphics[width=0.4\textwidth]{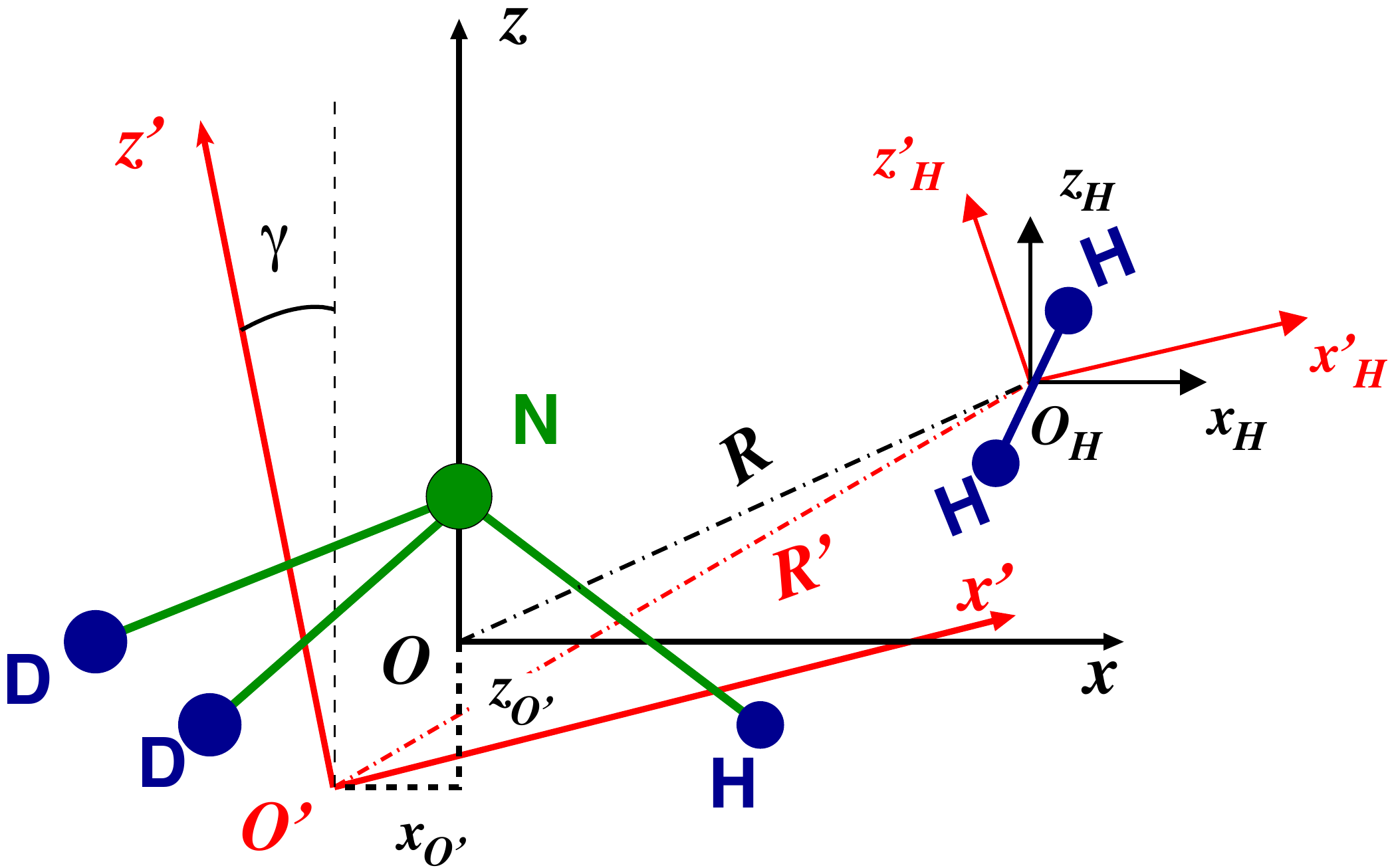}
\caption{ Scheme of the coordinate transformation between the frame
  $O'x'y'z'$ of ND$_2$H, and the original frame $Oxyz$, of the
  NH$_3$-H$_2$ system, see Eqs.\ (\protect\ref{eq:rototrasl}),
  (\protect\ref{eq:roto}). Only the $Oxz$ and $O'x'z'$ axes are
  represented, since the $y$ axes do not change. Distances are not to
  scale.}\label{fig:frames}
\end{figure}

The second point deals with the fitting of the \emph{ab initio} points
onto a functional suited for the scattering calculations. The PES
points resulting from the \emph{ab initio} procedure are described in
the following coordinate set: $R$, the distance between the centre of
mass of NH$_3$ and the centre of mass of H$_2$, $\theta, \phi$, the
corresponding spherical angles. For NH$_3$, $\theta$ is defined with
respect to $Oz$, the $C_{3v}$ axis of symmetry of NH$_3$, with the
nitrogen atom at positive values of $z$. The angle $\phi$ is the polar
coordinate in the perpendicular $Oxy$ plane, with one of the hydrogen
atoms of NH$_3$ lying in the $Ozx$ plane. Similarly, the H$_2$
molecule is oriented by the spherical angles $\theta_{\mathrm{H}},
\phi_{\mathrm{H}}$, in a reference frame
$O_{\mathrm{H}}x_{\mathrm{H}}y_{\mathrm{H}}z_{\mathrm{H}}$, parallel
to $Oxyz$ , with its origin at the centre of mass of the H$_2$
molecule (see Fig.~1). In order to define the new coordinate system
for ND$_2$H, we must first anchor the new centre of mass in the old
coordinate system, then perform the relevant rotation of the $Oxz$
axes around the invariant $Oy$ axis, which remains perpendicular to
the plane of symmetry of the ND$_2$H molecule. If $O'x'y'z'$ is the
new spherical frame centered on the $O'$ centre of mass on ND$_2$H,
and $\gamma$ the rotation of the $O'z'$ axis around the $O'y'$ axis
($O'y'$ is parallel to $Oy$), we have according to \citet{coh82}~:
\begin{eqnarray}
x_{O'}=-0.095957\,\text{Bohr} \nonumber \\ z_{O'}=-0.062152\,
\text{Bohr}\\ \gamma=11.30\,\text{degrees}\nonumber\label{eq:data}
\end{eqnarray}

It is enough to translate the centre of mass and perform the relevant
rotation $\gamma$ around $Oy$ and $O_{\mathrm{H}}y_{\mathrm{H}}$ to
get the new coordinate system in which the ND$_2$H-H$_2$ PES should be
expressed (see Fig.~\ref{fig:frames}) . The change of coordinates is
given by:

\begin{equation}
\left\{ \begin{array}{c}
  x=x'\cos\gamma-z'\sin\gamma+x_{O'}\\ y=y'\\ z=x'\sin\gamma+z'\cos\gamma+z_{O'}\end{array}\right.\label{eq:rototrasl}\end{equation}
The \hh ~ internal coordinates are transformed in a similar way, so
that $O'x'y'z'$ and
$O_{\mathrm{H}}x'_{\mathrm{H}}y'_{\mathrm{H}}z'_{\mathrm{H}}$
remain parallel and $O_H$ is now located with respect to $O'x'y'z'$ at
a distance $R'$. This corresponds to a pure rotation of $\gamma$ along
the $y_H$ axis:

\begin{equation}
\left\{ \begin{array}{c} x_H=x_H '\cos\gamma-z_H '\sin\gamma\\ y_H=y_H
  '\\ z_H=x_H '\sin\gamma+z_H
  '\cos\gamma\end{array}\right.\label{eq:roto}\end{equation}

Eventually, the functional form we take is identical to the one for
water-hydrogen \citep{phillips:94,valiron:08}. The spherical angles,
all primed, refer to the new frames $O'x'y'z'$ and $O_Hx'_Hy'_Hz'_H$ :
\begin{equation}
V\left(R',\theta',\phi',\theta'_{\mathrm{H}},\phi'_{\mathrm{H}}\right)=
\sum
v_{l_1l_2lm}(R')t_{l_1l_2lm}\left(\theta',\phi',\theta'_{\mathrm{H}},\phi'_{\mathrm{H}}\right)\label{eq:expansion},
\end{equation}
where the function $t_{l_1l_2lm}$ are explicitly given in
\cite{phillips:94,valiron:08}. As previously
\citep{valiron:08,maret:09}, we selected iteratively all statistically
significant terms $v_{l_1l_2lm}(R')$ using the procedure described in
\citet{valiron:08}. The final expansion includes anisotropies up to
$l_1$=10 for ND$_2$H and $l_2$=4 for H$_2$, resulting in 197
$t_{l_1l_2lm}$ functions.

%______________________________________________________________
%

\section{Collisional treatment}
\label{sec:collision}

In the low temperature environments where \nddh ~has been detected the
kinetic temperature is below 20~K \citep{roueff:00} and the main
constituent, \hh, is expected to lie in its ground rotational (para)
state, $J=0$ \citep{troscompt:09}. The calculations presented below
were restricted to collisions between ND$_2$H and para-H2($J$ = 0),
i.e. by neglecting the $J=2$ (closed) channel. This latter was
actually found to change the cross-sections, at very low energies, by
up to a factor of 3, as observed for NH$_3$-H$_2$ (see references in
\citet{maret:09}). Its effect on the average cross-sections and rate
coefficients is however much smaller, typically 30~\%, which is
therefore the typical accuracy of the rates below. The rotational
constants of ND$_2$H were taken from \citet{cou86}. The resulting rotational
energy levels in are given  Table \ref{tab:levels}. The reduced collisional mass is
$\mu=1.822684$~amu.

Full quantum close coupling (CC) scattering calculations were
performed with help of the Molscat program \citep{mols} which computes
scattering matrices ($S$ matrices) and combines these in order to
determine elastic and inelastic cross sections, at defined total
energies $E$. The convergence of the inelastic cross sections was
checked to be better than 5\%, for the different rotational
transitions investigated here. The different collision parameters
defined in Molscat are given the default ones. We used the following values : $\text{RMIN} =3.0$,
$ \text{INTFLG} =6 $ , $\text{STEPS}= 10.0-30.0$. Calculations
have been performed for total energies between 10~cm$^{-1}$ and
430~cm$^{-1}$. The energy step has been varied with increasing
collision energy. It was  0.25 cm$^{-1}$ for a
total energy between 10 cm$^{-1}$ and 40 cm$^{-1}$, 1 cm$^{-1}$
between 40 cm$^{-1}$ and 80 cm$^{-1}$, 5 cm$^{-1}$ between 80
cm$^{-1}$ and 100 cm$^{-1}$ and 20 cm$^{-1}$ above 100 cm$^{-1}$.
The size of the rotational basis set of ND$_2$H was also varied with
energy. For $E\leq 100\, \rm cm^{-1}$, $J(\mathrm{ND_2H})\leq 10$; 
at $E=120\, \rm cm^{-1}$, $J(\mathrm{ND_2H})\leq 12$. For the CS calculations at 
$E\geq 150\, \rm cm^{-1}$, $J(\mathrm{ND_2H})\leq 16$.

In order to limit the time of computation needed for a scattering
calculation, some customary approximations of the collisional
treatment are implemented in the Molscat program such as the Coupled
State (CS) approximation introduced by \citet{mcg74}. In this
approximation, scattering equations are written in the body-fixed
frame which is rotating. Then the projection of the orbital angular
momentum $l$ is restricted to the value $m_l =0$. The accuracy of this
method is questionable but it was shown to agree with CC calculations
within typically $\sim$~30\% for the present system, as shown
below. Here we resorted to the CS approximation for total energies
$E\geq 150\,\rm cm^{-1}$. As a result, at the
low temperatures at which the rates were calculated ($T\leq 35$~K),
the sensitivity to the CS calculations was checked to be marginal.

%--------------------------------------------------------

%
\begin{table}
\caption{Rotational level energies for ND$_2$H, as used in our
  computation. }
\label{tab:levels}
\begin{center}
\begin{tabular}{|l|l|}
\hline\hline $J_{K_aK_c}$& Energy (cm$^{-1}$)\\ \hline $0_{00}$ & 0\\
\\ $1_{01} $ & 9.09393 \\ $1_{11} $ & 11.19743 \\ $1_{10} $ & 12.78456 \\
\\ $2_{02} $ & 26.65954 \\ $2_{12} $ &
27.79543\\ $2_{11} $ & 32.55350  \\ $2_{21} $ & 38.86400 \\ $2_{20} $ & 39.48154
\\ \hline\hline
\end{tabular}
\end{center}
\end{table}%
%______________________________________________________________
\section{Results}
\label{sec:result}

\begin{figure*}
\includegraphics[width=0.9\textwidth]{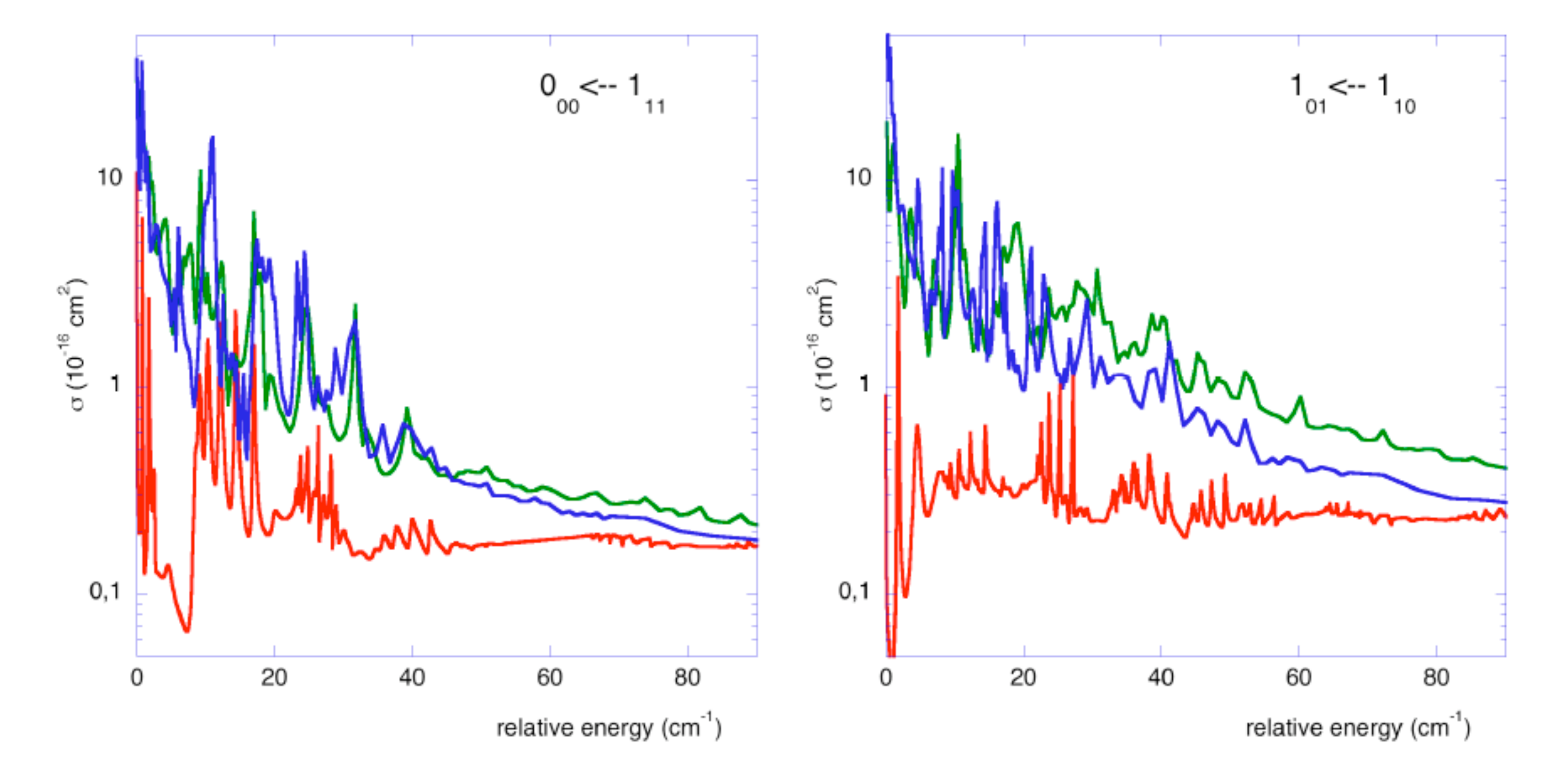}
\caption{ Collisional de-excitation cross sections as a function of
  the relative kinetic energy. Left: transition $1_{11} \rightarrow
  0_{00}$ of ND$_2$H with Helium in CS approximation (red)
  (\protect\citep{machin:07}) and para H$_2$ (present work) in CC
  (blue) and CS (green) approximations. Right : transition $1_{10}
  \rightarrow 1_{01}$ of ND$_2$H with Helium in CS approximation
  (red)(\protect\citep{machin:07}) and para H$_2$ (present work) in CC
  (blue) and CS (green) approximations}\label{fig:sec1}
\end{figure*}

Figure~\ref{fig:sec1} displays the collisional de-excitation cross
sections corresponding to the transition $1_{11} \to 0_{00}$ and
$1_{10} \to 1_{01}$ as a function of the relative kinetic energy
between ND$_2$H and H$_2$, both at the CC and CS level. These
transitions are observable from the ground at 335 and 110 GHz and are
potentially important probes of cold prestellar cores
\citep{lis:06,gerin:06,roueff:00}. For comparison, also cross sections
with He computed at the CS level, from \citet{machin:07}, are shown. A
complex resonance structure is seen for energies smaller than about
100 cm$^{-1}$ with broad and narrow features, resulting in large
variations of the cross sections.  The sharp maxima are due to the
opening of new collision channels and correspond to both Feshbach and shape
resonances. This is in contrast with the He-NHD$_2$ collision case, with a much shallower potential well. Even if
the symmetry of the two problems are identical, differences arise in their cross-section structures and cross section values. A similar effect was found recently at much higher energies, for the differential inelastic cross section
 of water-H$_2$, $J=0$ compared with water-He \citep{yang:10}. As is always been observed, the  resonances quickly disappear with increasing energy. The cross
sections relative to Helium are significantly smaller in the energy
range displayed in Figure~\ref{fig:sec1}. However, they converge to
those involving para-H$_2$ at collision energies above
$\sim$80~cm$^{-1 }$. 

Knowing the importance of the hyperfine quadrupolar effect for $^{14}$N, nuclear spin $I=1$, extending those cross section calculations to include explicitely the hyperfine structure could be of interest. Hyperfine structure has indeed been observed with some precision in the lower transitions \citep{gerin:06}. However, because of the low symmetry of NHD$_2$ (asymmetric top), no close formula exist in the literature that could connect state to state hyperfine rates to state to state global rotational rates. One should resort to a full sumation of $S$ matrix elements in order to get all detailed hyperfine rates. Also, because of the small column density of NHD$_2$, the differential effects on the various hyperfine lines should remain small \citep{daniel:06}. 
%
%\subsection{Collision  rates}
%
\begin{table*}  \centering %  Checked LW Jul 2010
  \caption{Deexcitation rate coefficients  for ND$_2$H-para \hh\ ($J=0$) collisions. Fits to formula 
  (\protect\ref{eq:fit}) \ are given in the last 4 columns.}
  \label{tab:rate} 

  \begin{tabular}{l l | l l l l l l| c l l l l}
    \hline\hline
  & \multicolumn{7}{ c }{Rate coefficients $R_{if}$ ( $10^{-11}$ cm$^3$.s$^{-1}$)}& & \multicolumn{4}{c}{Fit coefficients} \\
  Temperature
   & 5K  & 10K & 15K & 20K & 25K & 30K & 35K & $a_n$& 0 & 1 & 2 & 3          \\
\hline
$1_{01} \to 0_{00}$  & 0.572 &  0.436 &0.351  & 0.29  & 0.245 & 0.211  &0.183& &19.71 & -46.57 & 37.81 & -10.74 \\
 \\
$1_{11} \to 0_{00}$ &   1.14  &1.10  & 1.03  &0.938 &0. 854  & 0.781 & 0.717& & 22.53 & -51.13 & 38.73 & -9.74     \\
$1_{11} \to 1_{01}$ &   7.57 &6.79& 6.14 &5.62& 5.21 &4.89 & 4.63 &  &0.00 & -3.51 & 5.63 & -1.38    \\
 \\
$1_{10} \to 0_{00}$ &  4.75 & 4.39 & 4.05 & 3.74 & 3.49 &3.29&  3.13 & &
  1.89 & -7.36 & 8.07 & -2.04 \\
$1_{10} \to 1_{01}$ & 1.49&   1.34 &  1.23  &1.12 &1.03 & 0.953& 0.887 & &
 15.74 & -35.56 & 27.24 & -6.91 \\
$1_{10} \to 1_{11}$ &  1.37&1.25  & 1.14&  1.05 &  0.959 &  0.887&  0.826 & &
 14.76 & -33.90 & 26.31 & -6.76   \\
\\
$2_{02} \to 0_{00}$ &  0.824 &   0.782 &  0.71 &   0.64 &  0.581  & 0.532 & 0.493 &
& 6.58 & -19.33 & 17.89 & -5.40  \\
$2_{02} \to 1_{01}$ &  1.13 & 0.842 &  0.709 & 0.613 & 0.537 & 0.476 & 0.425 &
& 30.13 & -63.93 & 46.67 & -11.72  \\
$2_{02} \to 1_{11}$ &  3.19 & 3.06 &2.97 & 2.90& 2.83  &2.77 & 2.73 &
& 2.52 & -5.88 & 4.76 & -0.82  \\
$2_{02} \to 1_{10}$ &   1.29 &  1.12 & 1.00 & 0.904 & 0.821 & 0.752 & 0.694  &
& 15.98 & -36.13 & 27.88 & -7.23  \\
\\
$2_{12} \to 0_{00}$ & 0.275   & 0.272 & 0.234   & 0.199 &  0.171& 0.148 &0.13 & 
& 6.25 & -24.01 & 24.97 & -8.41  \\
 $2_{12} \to 1_{01}$ &  3.95& 3.84 & 3.73  & 3.61& 3.52& 3.44 & 3.37  &
& 1.09 & -3.63 & 3.67 & -0.58  \\
 $2_{12} \to 1_{11}$ &  0.808 &0.868 &0.823 & 0.753 & 0.684 & 0.623 & 0.569& 
& 15.88 & -40.31 & 32.97 & -8.84  \\
 $2_{12} \to 1_{10}$ & 0.739 &0.787 & 0.744 & 0.683 & 0.621& 0.565 & 0.516 &
& 17.41 & -43.09 & 34.66 & -9.22  \\
 $2_{12} \to 2_{02}$ &  3.40 & 3.77 & 3.97 & 4.03& 4.03 & 4.01 & 3.99 &
& 11.93 & -25.60 & 17.74 & -3.40  \\
 \\
 $2_{11} \to 0_{00}$ & 0.104 & 0.116 &  0.102 & 0.0875&  0.0749 &0.0647  &0.0564 &   
& 2.64 & -19.41 & 23.36 & -8.68  \\
 $2_{11} \to 1_{01}$ &   1.95 &2.22 & 2.36 &2.40 & 2.40 &2.38 & 2.35 &
& 16.14 & -34.63 & 24.05 & -5.07  \\
 $2_{11} \to 1_{11}$ &   3.72& 4.02& 4.14& 4.14  & 4.08 &3.98 & 3.88  &
& 18.59 & -39.65 & 27.74 & -5.77  \\
 $2_{11} \to 1_{10}$& 0.464 & 0.489 & 0.464 & 0.428 & 0.393  &0.362 & 0.336 &  
& 10.75 & -28.86 & 24.57 & -7.06  \\
 $2_{11} \to 2_{02}$ & 2.78 & 2.99 & 3.00 & 2.96 & 2.90 & 2.85 & 2.81  &
& -1.02 & -0.82 & 2.40 & -0.45  \\
 $2_{11} \to 2_{12}$ &   1.08 &1.25 &1.23 &1.15 &  1.07 & 1.00 & 0.943 &
& 2.38 & -13.06 & 14.38 & -4.38  \\
 \\
$2_{21} \to 0_{00}$ & 0.341 & 0.346 & 0.342 & 0.334 & 0.325  & 0.317 & 0.311& 
& 4.09 & -10.33 & 8.44 & -2.71  \\
$2_{21} \to 1_{01}$ &   3.41& 3.86& 3.98 &3.93 & 3.82  &3.69 & 3.57  &
& 15.47 & -35.32 & 26.13 & -5.71  \\
$2_{21} \to 1_{11}$ &   2.21 & 2.33 & 2.43 &2.48 & 2.50 & 2.50 & 2.50  &
& 13.36 & -26.93 & 17.69 & -3.41  \\
$2_{21} \to 1_{10}$ &  0.616 & 0.593 & 0.577 & 0.554 & 0.529 & 0.506 & 0.486&   
& 17.15 & -36.63 & 26.22 & -6.52  \\
$2_{21} \to 2_{02}$ &1.55 & 1.34 &  1.20 &1.10 &1.01 &0.947 & 0.896 &
& 3.52 & -9.97 & 9.64 & -2.92  \\
$2_{21} \to 2_{12}$ &   3.50 & 3.02 & 2.75 &2.56 & 2.41  &2.29 & 2.19 & 
& 4.43 & -10.36 & 8.80 & -2.11  \\
$2_{21} \to 2_{11}$ &1.93 &2.07& 2.10 & 2.09 & 2.05  & 2.00 & 1.96 & 
& 10.90 & -24.41 & 17.82 & -3.94  \\
 \\
$2_{20} \to 0_{00}$ & 2.43 &2.39 & 2.31 &2.20 &2.11 &2.02 & 1.95  &
& 6.54 & -16.01 & 12.96 & -3.09  \\
$2_{20} \to 1_{01}$ & 0.631 & 0.649 & 0.649 & 0.638 & 0.625  & 0.612 & 0.600  &
& 6.85 & -15.70 & 11.80 & -3.10  \\
$2_{20} \to 1_{11}$ &  0.441&  0.408 & 0.389 & 0.367 &0.346 & 0.327 & 0.309  &  
& 21.50 & -45.57 & 32.53 & -8.20  \\
$2_{20} \to 1_{10}$ &  2.98 & 3.00 & 3.03 & 3.04 & 3.03 &3.02 & 3.00  &
& 7.58 & -15.32 & 10.22 & -1.77  \\
$2_{20} \to 2_{02}$ & 1.86  & 1.66 & 1.56 &1.47 &1.40 & 1.33 &  1.27  &
& 14.77 & -31.02 & 22.26 & -5.22  \\
$2_{20} \to 2_{12}$ &  1.09 &1.02 & 0.998 &0.97 &0.938 &0.906 & 0.876  &
& 20.42 & -41.66 & 28.53 & -6.55  \\
$2_{20} \to 2_{11}$ & 0.897 & 0.992 & 1.03  & 1.03 & 1.02 &1.00 & 0.989  &
& 10.70 & -24.06 & 17.45 & -4.11  \\
$2_{20} \to 2_{21}$ &0.405 &0.379 &0.365 & 0.346 & 0.324 & 0.303 & 0.284 &
& 32.14 & -67.28 & 47.18 & -11.49  \\
    \hline \hline
  \end{tabular}
  %Values in parentheses refer to powers of ten
\end{table*}

\par The collisional (de)excitation rates $R$ as a function of
temperature $T$ were obtained by a Maxwell-Boltzmann averaging of the
cross sections times the relative velocity. In the equation below,
$E_{kin.}.$ is the relative kinetic energy:
\begin{equation}\label{eq:rates}
 R_{J_{K_aK_c} \to J'_{K'_aK'_c}}(T) = \left( \frac{8k_B T}{\pi \mu}
 \right)^\frac{1}{2} \left( \frac{1}{k_B T} \right)^2 \int_0^\infty
 \sigma_{J_{K_aK_c} \to J'_{K'_aK'_c}}(E_{kin})\,\exp \left
 (-\frac{E_{kin}}{k_BT} \right )\,E_{kin} \mathrm{d}E_{kin}
\end{equation}
 $k_B$ is the Boltzmann constant and $\mu$ is the reduced mass of the
ND$_2$H - H$_2$ system. Cross sections having been computed for
collision energies up to above 390~cm$^{-1}$, the rate coefficients
may be calculated very safely for temperatures between 5 K and 35 K
which are relevant for cold prestellar cores.
Table \ref{tab:rate} gives the de-excitation rate coefficients for the
9 first rotational levels as well as the fitting parameters allowing
to display the rate coefficients for any temperature between 5 and 35
K {\it via} the following expression \citep{faure:04}, with $R$ in
$10^{-11}$ cm$^3$.s$^{-1}$:
\begin{equation}
\log_{10}R(T)= \sum_0^3 a_n\,T^{-n/6}\label{eq:fit}
\end{equation}
All fits have been verified to be precise to better than 1~\%. It
should be noted that the following inelastic transitions are strictly
forbidden in the CS approximation (owing to spurious selection rules):
$1_{01} \to 0_{00}$ , $2_{12} \to 0_{00}$ and $2_{11} \to 0_{00}$. We
see that the corresponding CC rates are not negligible, although about
one order of magnitude smaller than the largest rates. There is
therefore no obvious collisional propensity rules for the presented
transitions. We note that since the dipole moment has two non-zero
projections ($\mu_b$ and $\mu_c$), the corresponding radiative
selection rules are $\Delta$J= 0, 1 and $\Delta$K$_a$ = $\pm$ 1,
$\Delta$K$_c$ = $\pm$1 for transitions along $\mu_b$ and $\Delta$K$_a$
=$\pm$1, $\Delta$K$_c$ = 0, $\pm$2 for transitions along $\mu_c$. The
Einstein coefficients have been computed by \citet{cou06} and recalled
in \citet{machin:07}.
%
%______________________________________________________________
\section{Discussion and conclusions}
\label{sec:conclusion}
The comparison of the present rates with the values obtained by
\citet{machin:07} where Helium is the collider is informative. The
collision rates obtained in the present work when para-H$_2$ (J=0) is
the perturber are typically one order of magnitude larger than those
obtained with Helium. The origin of this discrepancy lies in the
significantly different potential wells of the NH$_3$-H$_2$ and
NH$_3$-He PES, as discussed by \citet{maret:09}. The usual scaling
between the collision rates obtained from the ratio of the reduced
masses is clearly not accurate and the calculations have to be
performed explicitly in order to obtain the appropriate rates with
molecular hydrogen. We display in Table \ref{tab:critdens} the
resulting critical densities for the para and ortho transitions. These
were obtained by simply dividing the Einstein coefficient of a given
transition by the corresponding de-excitation rate coefficient at
10~K. Note that other definitions are possible \citep{maret:09}.
   \begin{table}
  \caption{Calculated critical densities for the first {\it ortho} and
    {\it para} transitions of ND$_2$H with H$_2$ at 10 K.}
  \label{tab:critdens}
  \centering
  \begin{tabular}{l c c}
    \hline\hline
Transition & \multicolumn{2}{c}{Critical density (cm$^{-3}$)} \\
 & {\it ortho} & {\it para} \\
\hline
$1_{11} \to 0_{00}$ & 1.17 10$^6$ & 1.34 10$^6$ \\
$1_{11} \to 1_{01}$ & 3.48 10$^4$ & 6.63 10$^4$ \\
$1_{10} \to 0_{00}$ & 1.10 10$^7$ & 1.01 10$^7$ \\
$1_{10} \to 1_{01}$ & 5.19 10$^4$ & 5.94 10$^4$ \\
$2_{02} \to 1_{11}$ & 6.34 10$^5$ & 6.76 10$^5$ \\
$2_{02} \to 1_{10}$ & 5.33 10$^6$ & 5.66 10$^6$ \\
$2_{12} \to 2_{02}$ & 2.31 10$^4$ & 9.26 10$^3$ \\
$2_{11} \to 1_{01}$ & 1.14 10$^8$ & 1.18 10$^8$ \\
$2_{11} \to 1_{10}$ & 1.33 10$^7$ & 1.45 10$^7$ \\
$2_{11} \to 2_{02}$ & 7.59 10$^4$ &7.12 10$^4$ \\
$2_{21} \to 1_{11}$ & 1.80 10$^8$ & 1.79 10$^8$ \\
$2_{21} \to 1_{10}$ & 4.17 10$^7$ & 1.77 10$^7$ \\
$2_{21} \to 2_{12}$ & 2.70 10$^5$ & 1.94 10$^5$ \\
$2_{21} \to 2_{11}$ & 1.39 10$^6$ & 1.23 10$^6$ \\
$2_{20} \to 1_{11}$ & 5.78 10$^7$ & 2.26 10$^7$ \\
$2_{20} \to 1_{10}$ & 1.47 10$^8$ & 1.49 10$^8$ \\
$2_{20} \to 2_{12}$ & 5.65 10$^6$ & 6.44 10$^6$ \\
$2_{20} \to 2_{11}$ & 3.96 10$^5$ & 2.48 10$^5$ \\
 \hline \hline
  \end{tabular}
\end{table}
These values set the limit of the perturber density above which the
transitions become thermalized. We see that the critical densities
corresponding to the two transitions $1_{11} \to 0_{00}$ and $1_{10}
\to 0_{00}$ are different by about one order of magnitude at 10~K,
which is about a factor of 3 less than the ratio of the Einstein
coefficients. Such differences may have profound consequences on the
interpretation of these two transitions which are both observable from
the ground \citep{lis:06,gerin:06} at 335 and 389 GHz respectively, as
the corresponding upper levels may not be accounted for by a single
excitation temperature.

\section*{Acknowledgements} 
We are deeply indebted to Pierre Valiron, who gave impetus and ideas
throughout the inception of this work. ES was supported by a grant of
the "Molecular Universe" FP6 Marie Curie network. We acknowledge
support from the French Institut National des Sciences de l'Univers,
through its program ``Physique et Chimie du Milieu
Interstellaire''. LAOG is a joint laboratorory Universit\'e
Joseph-Fourier/CNRS under the name UMR 5571.

\bibliographystyle{mn2e}

\end{document}